\documentclass[prl,twocolumn,superscriptaddress]{revtex4-1}
\usepackage{graphicx}
\usepackage{amsmath}
\usepackage{dcolumn}
\usepackage{color}
\usepackage{epstopdf}
\usepackage[caption=false]{subfig}

\usepackage{float} 

\begin{document} 

\title{Nonlinear reconnecting edge localized modes in tokamaks}
\author{F. Ebrahimi}  
\affiliation{Princeton Plasma Physics Laboratory, and Department of Astrophysical Sciences, Princeton University NJ, 08544}

\date{\today}
 
\begin{abstract}
Nonlinear edge localized modes in a tokamak are examined using global three-dimensional resistive magnetohydrodynamics simulations. Coherent current-carrying filament (ribbon-like) structures wrapped around the torus are nonlinearly formed due to nonaxisymmetric reconnecting current sheet instabilities, the so called peeling-like edge localized modes. These fast growing modes saturate by breaking axisymmetric current layers isolated near the plasma edge and go through repetitive relaxation cycles by expelling current radially outward  and relaxing it back. The local 
bi-directional fluctuation-induced electromotive force (emf) from the edge localized modes, the dynamo action, relaxes the axisymmetric current density and forms current holes near the edge.

\end{abstract}

\maketitle

Magnetohydrodynamic (MHD) stability of the plasma boundary is critical for the successful performance of future magnetic fusion devices,
such as ITER.~\cite{iter} Quasiperiodic burst-like edge MHD instabilities observed routinely during the high performance operation regime of tokamaks, the H-mode,~\cite{h-mode} have been shown to degrade the global confinement.
Due to formation of a transport barrier during H-mode operation, steep pressure gradients and large currents exist at the plasma edge, which could provide free energy  for the so called Edge Localized Modes (ELMs).~\cite{elm_connor}. 
The quasiperiodic bursts of ELMs lead to loss of energy and particles on a short time scales of 0.1-1~ms and could 
potentially damage vessel
components due to high local heat fluxes. ELMs are known to be ideal-like MHD modes of coupled peeling (current-driven) and ballooning (pressure-driven) instabilities.~\cite{snyder2002,wilson2002} Ideal (i. e. non-resistive) MHD codes, as well as extended MHD codes,~\cite{ferraro2010,burke} have been extensively used for linear studies of these modes. However, the effect of collisions as well as \textit{full} nonlinear 3-D MHD dynamics of these modes have not yet been explored. In particular the physics of their repetitive cycles has not been understood until now. 

In this Letter, we explore the nonlinear dynamics of current-driven ELMs using extended MHD simulations in a toroidal tokamak configuration.  Understanding the physical dynamics, 
as well as poloidal/toroidal localization and the radial extent of these structures, is crucial for mitigating their impact. Here, for the first time, coherent current carrying filament edge localized structures with repetitive nonlinear dynamics are demonstrated during nonlinear MHD resistive simulations in a spherical tokamak. First, we perform simulations with varying toroidal magnetic field ($B_{\phi}$), but keeping the 
current-density profile fixed, and find that the growth of current-driven reconnecting edge localized modes (with tearing parity) scales with $J_{||}/B$ and 
stabilization occurs at high $B_{\phi}$. Second, we then show that these reconnecting edge modes with low toroidal mode numbers ($n=1-5$) saturate and form nonaxisymmetric filament current-carrying  structures during the nonlinear stage. These nonlinear structures go through quasiperiodic oscillations. Third, we explain the physics of the nonlinear dynamics of the  
field-aligned filaments via direct calculations of the fluctuation induced emf term, the vector product of flow and magnetic field fluctuations.  The emf terms are localized around the edge current layers, where the reconnecting edge modes are triggered. It is also 
found that the current-driven edge modes alone (an important component of ELMs), without the pressure-driven component,  are sufficient to explain 
  the quasiperiod cycles along with the rapid relaxation of edge current with a time scale of the order of 0.1ms. The coherent filament structures found here are in particular very similar to the experimentally observed current-carrying filaments from peeling modes in Pegasus~\cite{2011peeling}. Filament structures during ELMs have been experimentally observed in other low-aspect ratio spherical tokamaks, MAST~\cite{kirk_mast} and NSTX~\cite{maingi}.

Edge localized peeling modes are triggered due to strong edge currents. In tokamaks, 
finite edge current density typically exists either due to finite edge temperate (allowing Ohmic current flows) or finite edge pressure gradient (allowing bootstrap currents).~\cite{elm_connor}. In this study, the edge currents are formed during the plasma startup current drive using the electrostatic helicity injection technique called Coaxial Helicity
Injection (CHI), the primary candidate for non-inductive plasma
current start-up in NSTX-U.~\cite{ebrahimi16}
 We show that the driven current structure at the edge of the injected flux could trigger current-driven edge localized modes.  Plasma generated during helicity injection current start up provides a unique 
platform to study nonlinear evolution of edge localized 
current-driven instabilities in tokamaks in isolation, without the influence of other types of instabilities (pressure-driven ballooning).

\begin{figure}[!b]
\includegraphics[]{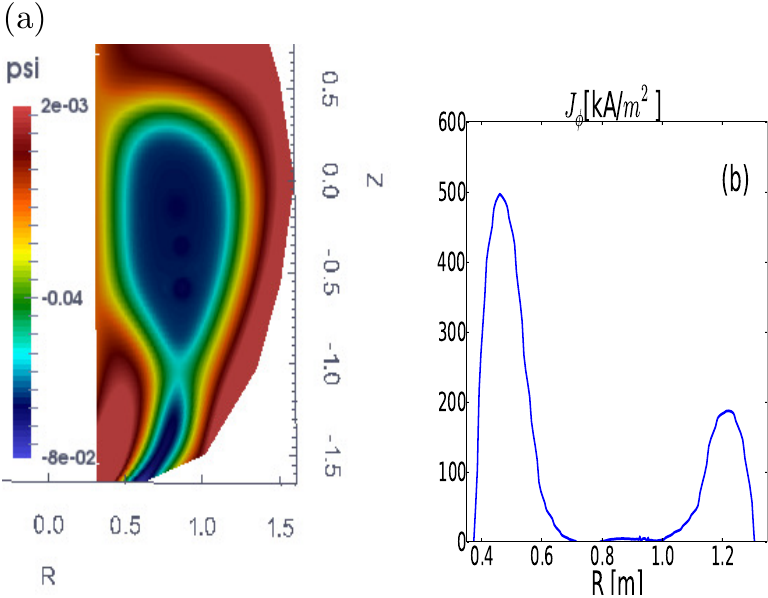}
\caption{Typical poloidal flux (Wb) and axisymmetric toroidal current density, $J_{\phi} \, (A/m^2)$ during nonlinear simulations. }
  \label{fig:fig1} 
\end{figure}

The startup plasma current is formed by injecting biased poloidal flux, i.e. by driving current along the injected open field lines (the injector current) 
in the presence of an external toroidal field.(for details see Fig. 1 in~\cite{ebrahimi2015plasmoids})  Helicity injection for current statup  is initiated by applying a voltage, or a constant electric field, to the divertor plates (the poloidal flux
footprints). Helicity is injected through the linkage of resulting toroidal flux with the poloidal injector flux. Plasma and open field lines (the magnetic bubble) expand into
the vessel if the injector current exceeds a threshold value. Figure ~\ref{fig:fig1}(a) shows the
typical expanded poloidal flux.  As the poloidal flux is expanded in the volume, an edge current sheet at the edge of the injected flux is formed (Fig.~\ref{fig:fig1}). This edge current density layer is formed on both high and low field sides of the tokamak as seen in Fig.~\ref{fig:fig1}(b). The current profile, although hollow in the core during the CHI, does have a 
peak near the edge, which could provide the free energy for nonaxisymmetric edge instabilities studied here. The current density spikes generated non-inductively here (Fig.~\ref{fig:fig1}(b)) are similar to the edge current spikes during ELMs in tokamaks.

To examine the nonlinear evolution of current-driven edge localized instabilities, we perform nonlinear MHD simulations using the NIMROD code.~\cite{sovinec04}
We use a poloidal grid with 45 $\times$
90 fifth-order finite elements in a global (R,Z) geometry,
and toroidal Fourier mode numbers $n\neq 0$ up to 43 in 3-D simulations. A
uniform number density of $4 \times 10^{18}m^{-3}$ for a deuterium
plasma is used. The helicity injection model, boundary condition and NSTX/NSTX-U geometries are the same as in earlier papers~\cite{ebrahimi2013,hooper2013,ebrahimi16}. To isolate the effect of current-driven instabilities localized near the edge, we perform resistive MHD simulations for zero pressure model (pressure is not evolved
in time). Any edge pressure-driven instabilities, including ballooning modes, are  therefore eliminated here.  We used magnetic diffusivitites in
the range of $\eta=2.5-12 m^2/s$ to give Lundquist
number in the range of $S=L V_A/\eta=1-5\times 10^5$.  Here,
the Alfven velocity  $V_A$ is based on the reconnecting 
magnetic field,  $L$ is the current sheet length. The kinematic viscosities are chosen to give a Pm =7.5 (Prandtl number= $\nu/\eta$)

\begin{figure}
\includegraphics[]{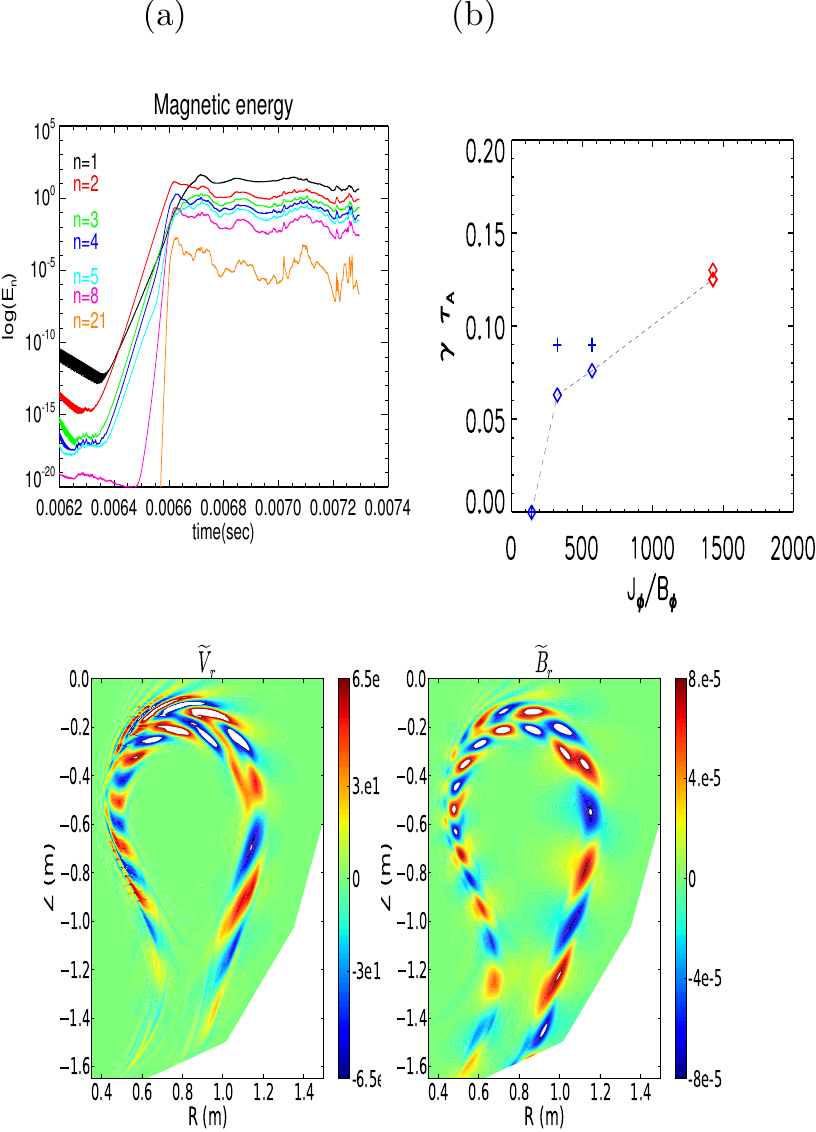}
\caption{Top left: Total magnetic energies of different toroidal mode numbers vs. time during nonlinear 3-D simulations ($\eta=5 m^2/s$, $B_{\phi}=0.7T$, peaked current density near the edge $J_{\phi}$ =500 kA/$m^2$ and reconnecting field $B_z=0.1T$). Top right: Growth rates ($\gamma \tau_A$) of n=1 modes vs. $J_\phi/B_\phi[kA/(m^2.T)$]. The tree blue diamond (cross) points are n=1 (n=2) growth rates when peak current density is fixed ($J_\phi$ =400kA/$m^2$) and only toroidal field is varied ($B_\phi$ = 2.8, 1.23, and 0.7T) (S=11000). Red diamonds are n=1 growth rates for $J_{\phi}=1. MA/m^2$ and $B_{\phi}=0.7$ ($\eta=2.5,8 m^2/s$).
Bottom: typical linear mode structures, radial velocity and magnetic perturbations, of edge localized n=1 mode.}
  \label{fig:fig2} 
\end{figure} 

We first perform three-dimensional simulations with the initial poloidal flux (and the associated current density) shown in Fig.~\ref{fig:fig1}. The peak of toroidal current density 
in the current layer is about 500 kA/$m^2$ in the region of high toroidal field side ($B_\phi$ =0.7T) as shown in Fig.~\ref{fig:fig1}(b). In 3-D, non-axisymmetric 
magnetic fluctuations arise due to current-sheet instabilities localized near the edge region. Figure~\ref{fig:fig2} shows the magnetic energy of different toroidal mode numbers during nonlinear evolution. As it is seen, magnetic fluctuations with low toroidal mode numbers n=1-6 are linearly unstable and saturate, while fluctuations with higher toroidal modes numbers grow only nonlinearly and saturate at much lower  amplitudes.  The mode with longest wavelength, n=1, grow with growth rate of $\gamma \tau_A =0.086$, and also n=2-6 grow fast. Here, $\tau_A$ is the poloidal transit time based on the 
local reconnecting field $B_z\approx 0.1T$. The linear mode structure of the  nonaxisymmetric mode, n=1, is shown in Fig.~\ref{fig:fig2}. This mode is localized around the edge current sheet (the edge region with the current density spike) and has a tearing parity (radial magnetic and velocity magnetic 
fluctuations are even and odd around the current layer, respectively).~\cite{ebrahimi2016dynamo} The radial velocity changes sign in the current layer, similar to peeling mode structures with tearing parity observed near the X point region reported in~\cite{2005peeling}. As it is seen this mode has a high poloidal mode (m) number. Considering periodicity in the poloidal direction, this is expected in the edge region with high winding number $q=rB_{\phi}/RB_p=m/n$. A similar current-driven n=1 mode with high poloidal mode number  structure was also triggered in simulations with different initial conditions for a specific case of CHI in NSTX.~\cite{hooper2016}

Next, to further investigate the nature of these edge localized modes, we 
 perform several simulations by varying the toroidal field, but with the same current density profile. We calculate the growth rates of these modes during the early
linear phase of these nonlinear simulations. The first three points in Fig.~\ref{fig:fig2}(b)
 show the growth rates for simulations with three different toroidal fields. The peak toroidal 
current density (and its profile) in the edge current layer kept the same ($J_\phi$ =400kA/$m^2$). At high toroidal guide field and low values of $J_{\phi}/B_{\phi} $, the edge localized current-driven n=1 modes are stable. The instability grows fast at higher values of about $J_{\phi}/B_{\phi} >150-200$ as shown in Fig.~\ref{fig:fig2}(b). The scaling of the instability with  $J_{\phi}/B_{\phi} $ is consistent with the instability drive for the traditional ideal peeling modes, $qR J_{||}/B$.~\cite{elm_connor} 
However, here in the presence of resistivity which leads to a resolved physical current sheet, the criteria for the instability will be different from the ideal peeling mode stability criteria. In the presence of resistivity, physical current layers developing
 near the plasma edge can be unstable to current-sheet instability.~\cite{ebrahimi2016dynamo}
As the unstable mode here has reconnecting tearing parity, linear analysis for
 current-sheet instability such as oblique plasmoid instability ~\cite{baalrud} is therefore more relevant. Our limited scaling study for three S values in the range of ($1-4 \times 10^5$) show a week dependency on S. There is opportunity for further linear theory development for the tokamak edge current sheet equilibrium profiles,  for example incorporation of the edge current sheet width scaling with S into the tearing analysis for tokamak edge modes.

 \begin{figure}
\begin{tabular}{p{5.2cm}p{5.2cm}}
\includegraphics[width=3.2in,height=2.1in]{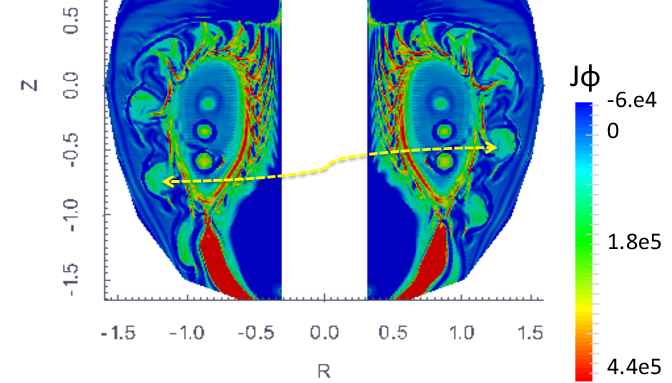}\\
\includegraphics[width=2.in,height=1.8in]{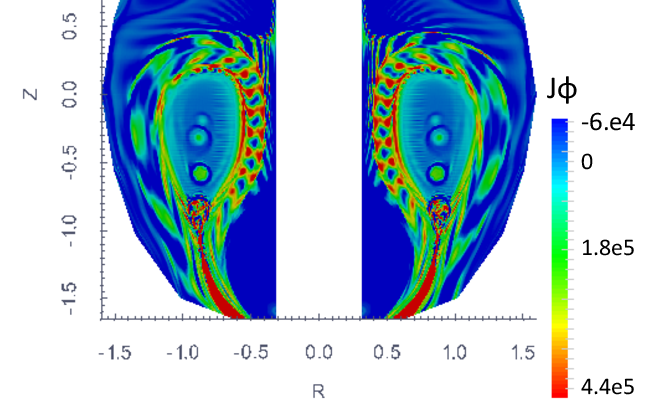}&
\includegraphics[width=1.2in,height=1.8in]{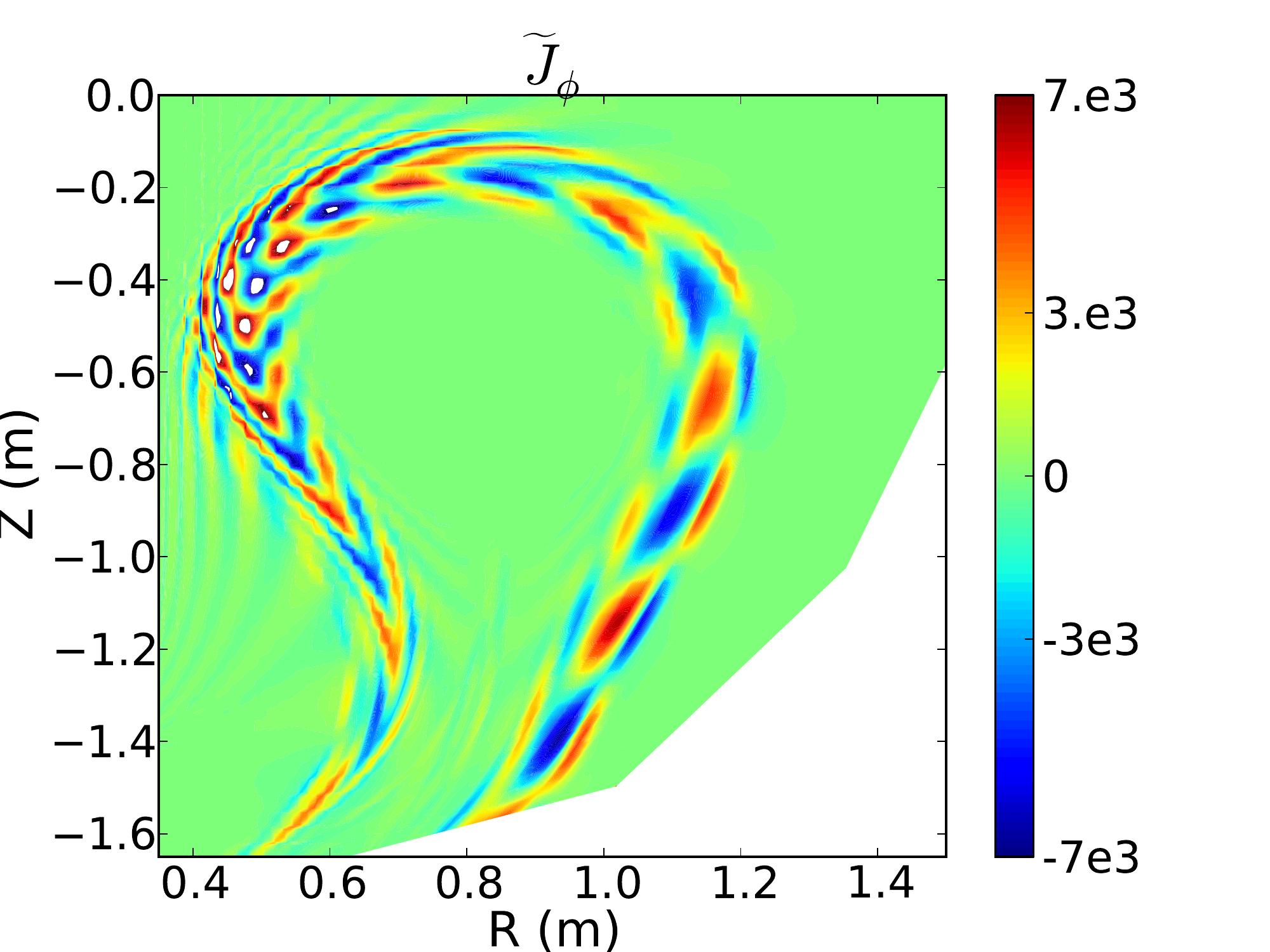}
\end{tabular}
\caption{Poloidal R-Z cuts of saturated nonlinear toroidal current density for the case shown in Fig.~\ref{fig:fig2}(a) at t=7ms, top: with 22 toroidal Fourier modes, bottom left: with only 2 toroidal Fourier modes. Bottom right: linear toroidal current density mode structure during the growth phase of n=1.} 
  \label{fig:fig3} 
\end{figure}

Following the linear stage studied above, the modes saturate as shown from the magnetic energy evolution 
 in Fig.~\ref{fig:fig2}(a).  Fig.~\ref{fig:fig3}  shows the saturated nonlinear current density for the case above (with energy shown in Fig.~\ref{fig:fig2}) at t=7ms. As seen, coherent non-axisymmetric n=1 structures have evolved during the nonlinear saturated state. These nonaxisymmetric structures break the initial axisymmetric current density (Fig.~\ref{fig:fig1})(b). The edge localized modes with low toroidal mode number saturate with magnetic perturbations amplitudes of about a few percent of n=0 reconnecting magnetic field (0.1T). However, because of the very localized nature of the magnetic fluctuations, the amplitude of the 
associated perturbed toroidal current density can be as high as 50\% of the n=0 component.
  For comparison, the nonlinear structure in simulations with only two toroidal Fourier modes and linear structure are shown in the Fig.~\ref{fig:fig3} (lower panel). The radially propagating filament n=1  structures can only become coherent as the number of toroidal modes increased as seen in Fig~\ref{fig:fig3}(a). The nonlinear filament 
structures are radially  extended about 30-35 cm (Fig.~\ref{fig:fig3}(a)), while the linear mode structure has only a radial extension of about 10-15cm (Fig.~\ref{fig:fig3}(c)).
These localized coherent nonaxisymmetric structures are also nonlinearly formed due to the strong edge current layer in other simulations with different initial poloidal flux, and higher peaked current density and S. The nonlinear dynamics of these structures for simulations at higher toroidal field will be addressed below.

\begin{figure}[!b]
\vspace{-12mm}
\begin{tabular}{p{4.2cm}p{4.2cm}}
\includegraphics[width=2.5in,height=2.1in,angle=180]{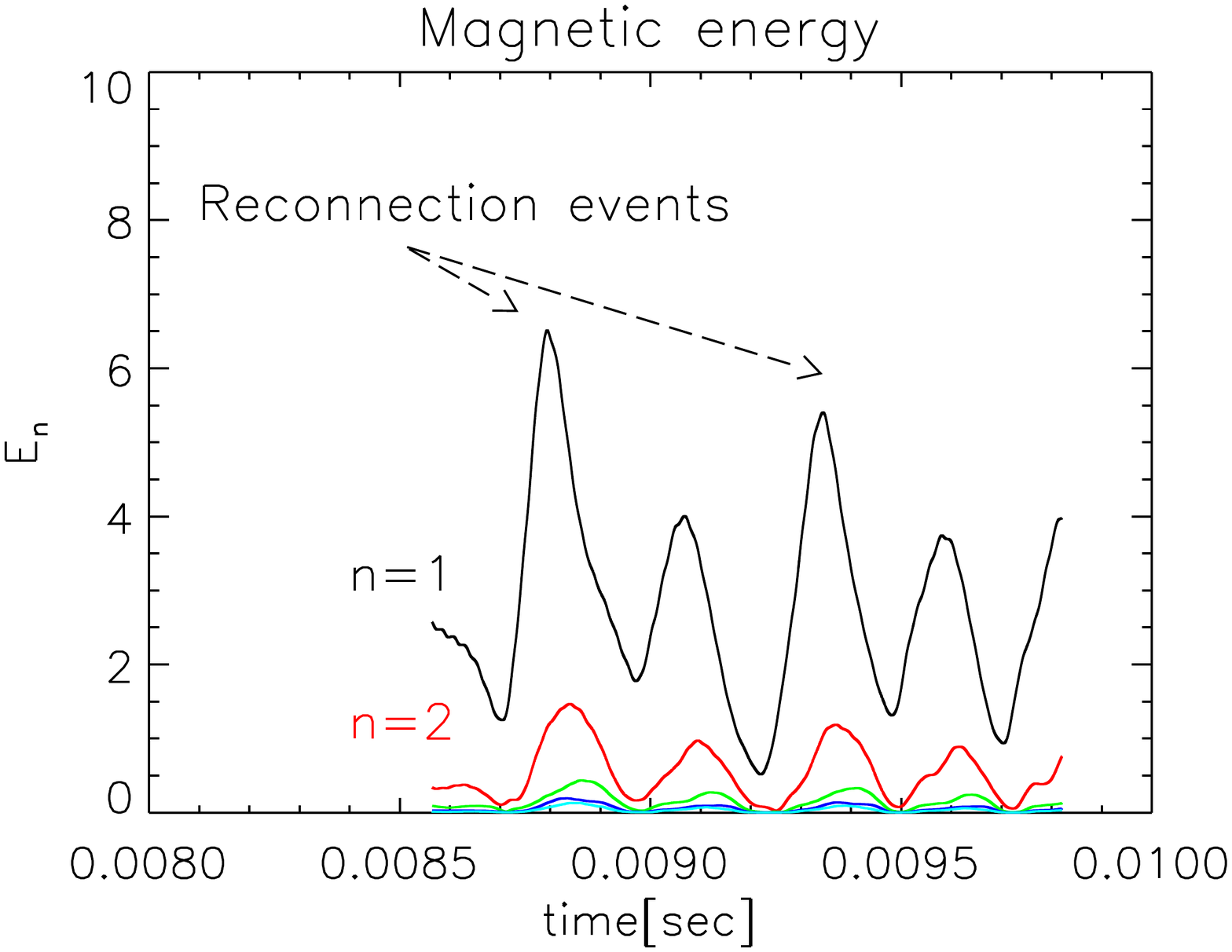}\\
\vspace{-3mm}
(b) \hspace{31mm} (c) \\
\includegraphics[width=1.65in,height=2.1in]{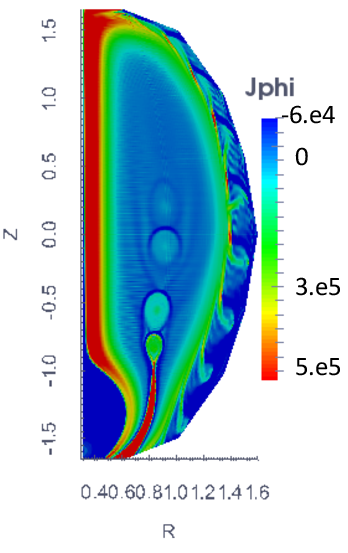}&
\includegraphics[width=1.65in,height=2.1in]{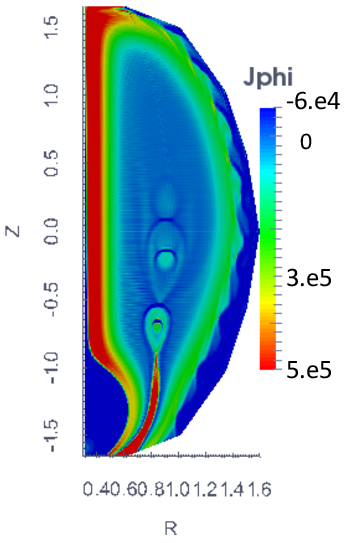}
\end{tabular}
\caption{Top: Cyclic oscillations of total magnetic energies of edge localized modes during nonlinear 3-D simulations in NSTX-U geometry with $B_{\phi}=1.23T$. Poloidal R-Z cuts of saturated nonlinear toroidal current density at t= 8.8ms (bottom left) and t= 8.7ms(bottom right). }
  \label{fig:fig4} 
\end{figure}

\begin{figure}
\includegraphics[width=3.2in]{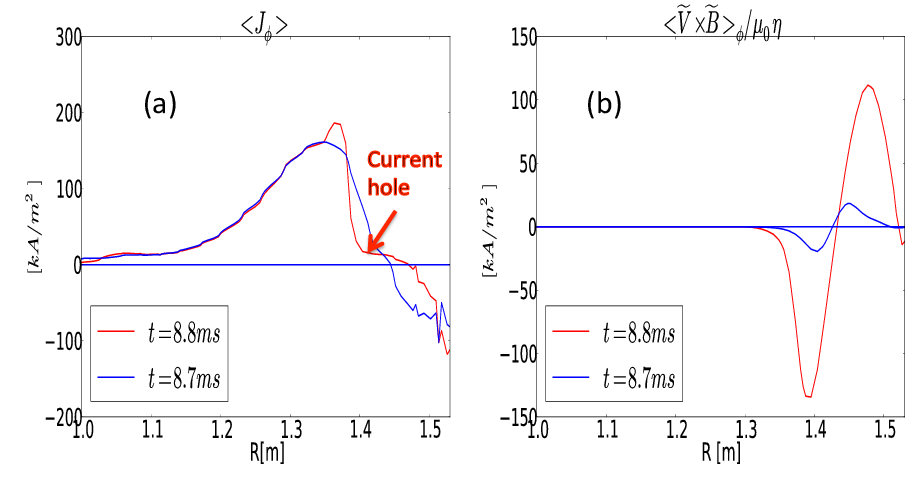}
\caption{Profiles of (a) $<J_{\phi}>$ averaged vertically (b) normalized (R,$\phi$)-averaged 
toroidal emf term $<\tilde{\textbf{V}} \times \tilde{\textbf{B}}>_\phi$, at two times during a cycle.}
  \label{fig:fig5} 
\end{figure}

Observed edge localized coherent structures exhibit repetitive cycles during nonlinear stage (top panel Fig.~\ref{fig:fig4}). We now analyze the behavior of nonlinear localized edge coherent structure during a cycle. Top panel of Fig.~\ref{fig:fig4} shows the total energy vs. time during full nonlinear 3-D simulations with 43 toroidal modes and with high toroidal field ($B_{\phi}$ =1.23T). We study the nonlinear
 dynamics during the flat top part of the total current (320kA). Axisymmetric edge toroidal current layers are formed poloidally (on both low field and high field sides). However, due to high toroidal field, only the outer edge (on the low field side) remains strongly perturbed (as shown in Fig.~\ref{fig:fig4}).    The lower panel of 
 Fig.~\ref{fig:fig4} shows the poloidal cross sections of nonlinear total toroidal current densities at two times, at maximum (t=8.8ms) and minimum (t=8.7ms) fluctuations, during a cycle.  At the time with maximum fluctuation level, edge localized modes evolve to  form coherent nonlinear filament structures at the low field side (lower panel Fig.~\ref{fig:fig4}). These localized current carrying structures with poloidally (twisting vertically) and  toroidally (n=1 wrapping around the torus) localization  are also extended radially. The structures do relax back radially to merge back into a axisymmetric toroidal current density as seen in Fig.~\ref{fig:fig4} (lower right panel) near the the edge region of low field side.  Note that the axisymmetric plasmoids generated during helicity injection region seen in Fig.~\ref{fig:fig4} have been studied elsewhere.~\cite{ebrahimi2016dynamo} Here the focus is the coherent \textit{edge} localized structures. 

We examine these relaxation events by analyzing the fluctuation induced emf term (the correlated flow and magnetic field fluctuations $\tilde{\textbf{V}} \times \tilde{\textbf{B}}$) from the edge localized modes themselves. The fluctuations can affect the total current in two ways, first the fluctuations can reach a finite amplitude, second through the mean emf dynamo term to modify the average current density. We separate the fields into mean and fluctuating components ($\textbf{J} =<\textbf{J}>_{n=0} +\tilde{\textbf{J}}_{n\neq 0}$), where mean $<>$ is vertically and toroidally (Z, $\phi$) averaged axisymmetric field. The coherent nonaxisymmetric structures with high poloidal mode numbers (as the edge localized modes themselves) create current holes in the current layer as seen in the low field side (Fig.~\ref{fig:fig4}(b)), and relax back (Fig.~\ref{fig:fig4}(c)). The modification of the edge current density and the formation of current holes can be explained through the mean fluctuation induced emf.

Figure~\ref{fig:fig5} shows the toroidal current density averaged vertically (averaged around mid plane $-0.5 m < Z < +0.5m$) at two times. During the low fluctuation part of the cycle (t=8.7ms) the current density remains mostly axisymmetric, with the positive spike of the current (radially symmetric around R=1.35m), in Fig.~\ref{fig:fig5}(a). 
However, the vertically-averaged current density is drastically different because of the nonaxisymmetric fluctuations at t= 8.8ms, and a current hole region around R= 1.38-1.45m is created (Fig.~\ref{fig:fig5}(a) red curve). The calculated toroidal fluctuation induced emf,  $<\tilde{\textbf{V}} \times \tilde{\textbf{B}}>_{\phi} = 1/2Re[\widetilde V_r \widetilde B_z^{*} - \widetilde V_z \widetilde B_r^{*}]$, 
 at these two times is also shown in Fig.~\ref{fig:fig5}(b). As seen, a nonzero fluctuation induced 
nonlinear dynamo term is bipolar and has a sufficiently large amplitude to contribute to the current density modification. The localized dynamo term changes sign around the same radius where the flattening and annihilation of current density occurs. It is therefore shown that the emf does contribute to the formation of current holes and the radially outward expulsion of the current density. It should be noted that the relaxation in a short time period of  0.1ms, shown in Figs.~\ref{fig:fig4},\ref{fig:fig5}, is consistent with experimentally observed time scales.

In conclusion, nonlinear coherent filament edge localized structures have been formed during the full nonlinear 3-D resistive MHD simulations in a tokamak configuration. The structures are nonaxisymmetric, poloidally localized, wrap around the torus, and radially extended. The mode structure in the closed flux region extends to the open field region during the nonlinear evolution. 
 For the first time, it has been shown that the edge localized structures in tokamaks 1) have reconnecting nature, 2) grow much faster (close to poloidal Alfven transit times) than the global tearing modes ~\cite{fkr} and 3) have finite localized fluctuation-induced bi-directional emf dynamo term ($<\tilde{\textbf{V}} \times \tilde{\textbf{B}}>$), which could cause the  local annihilation of axisymmetric current (current holes) near the edge region in the nonlinear stage. The coherent filament structures found here are very similar to the camera images of peeling modes from Pegasus spherical tokamak (see Fig. 1 in~\cite{2011peeling}).  Finally,  as solar eruptions are accompanied by the 
 ejection of field-aligned filamentary structures into the
surrounding space, the MHD study of nonlinear filaments 
here could also improve our understanding of these eruptive events on the sun.

\acknowledgements 
We acknowledge S. Prager for valuable comments on this Manuscript. Computations were performed at NERSC. This work was supported by DOE grants DE-SC0010565, DE-AC02-09CHI1466, and DE-SC0012467.

\end{document}